\documentclass[12pt]{article}
\usepackage{amsmath}
\usepackage{graphicx,psfrag,epsf}
\usepackage{enumerate}
\usepackage{natbib}
\usepackage{url} 
\usepackage[dvipsnames]{xcolor} 

\newcommand{\blind}{1}

\begin{document}

\def\spacingset#1{\renewcommand{\baselinestretch}%
{#1}\small\normalsize} \spacingset{1}

\newcommand{\shinyapp}[1]{\url{https://mdogucu.shinyapps.io/teachered-bayes-shiny/}}
\newcommand{\shinyrepo}[1]{\url{https://github.com/mdogucu/teachered-bayes-shiny}}
\newcommand{\googledoc}[1]{\url{https://docs.google.com/document/d/1G3H1aePCx7Jcg226v59m-LiTVcLYe1Ym/edit?usp=sharing&ouid=110969489876425298430&rtpof=true&sd=true}}
\newcommand{\slides}[1]{\url{https://docs.google.com/presentation/d/1L27DRrJFqTGGKv9gCgEYhvJfNntsPymH/edit?usp=sharing&ouid=116742544548104368401&rtpof=true&sd=true}}
\newcommand{\irb}[1]{University of Tennessee, Knoxville}


\if1\blind
{
  \title{\bf The Design and Implementation of a Bayesian Data Analysis Lesson for Pre-Service Mathematics and Science Teachers}
  \author{Mine Dogucu\thanks{
    The authors gratefully acknowledge Nick Kim for teaching the mathematics teaching methods class that partially served as the context for this study. Dogucu was supported by the National Science Foundation under Grant No. 2215879. Rosenberg was supported by the National Science Foundation under Grant No. 1937700.}\hspace{.2cm}\\
    Department of Statistical Science, University College London\\
    Department of Statistics, University California Irvine\\
    
    Sibel Kazak \\
    Department of Mathematics and Science Education, \\  Middle East Technical University  \\
    and \\
    Joshua Rosenberg \\
    Department of Theory and Practice in Teacher Education \\ 
    University of Tennessee, Knoxville}
  \maketitle
} \fi

\if0\blind
{
  \bigskip
  \bigskipƒ
  \bigskip
  \begin{center}
    {\LARGE\bf Title}
\end{center}
  \medskip
} \fi

\bigskip
\begin{abstract}

With the rise of the popularity of Bayesian methods and accessible computer software, teaching and learning about Bayesian methods are expanding. However, most educational opportunities are geared toward statistics and data science students and are less available in the broader STEM fields. In addition, there are fewer opportunities at the K-12 level. With the indirect aim of introducing Bayesian methods at the K-12 level, we have developed a Bayesian Data Analysis activity and implemented it with 35 mathematics and science pre-service teachers. In this manuscript, we describe the activity, the web app supporting the activity, and pre-service teachers' perceptions of the activity. Lastly, we discuss future directions for preparing K-12 teachers in teaching and learning about Bayesian methods.

\end{abstract}

\noindent%
{\it Keywords:}  

Bayesian methods, grades K-12, science education, mathematics education, teacher education

\vfill

\newpage
\spacingset{1.45} 

\section{Introduction}

\label{sec:intro}

One spring, students eating lunch at a high school may observe there are more birds visiting a courtyard outside their school than they noticed in the past. Why might this be? Students (and adults) may generate a range of plausible ideas. But, regardless of any specific hypotheses, any conclusions achieved will not be certain. Many things could cause greater visits from birds, and no answer or any single answer will offer the last word on this question.

Students encounter uncertainty in other parts of the school, too. Why was the traffic worse today than yesterday? How come an investigation during a Chemistry laboratory did not work out as well for students in one laboratory group as another? How confident about historical knowledge should one be after reviewing the histories of people living 100s or 1000s of years in the past? Answers to these questions are necessarily uncertain. 

While uncertainty is present in students' everyday experiences, how we teach about uncertainty in school often falls short of offering learners useful strategies or mathematical or statistical approaches that can bolster how they generate answers to real-world scientific questions. 




The challenge in understanding uncertainty is not unique to students. 
Many scientists (and statisticians) also struggle with uncertainty in answering scientific questions. If trained in statistics, scientists and students are more commonly trained in frequentist statistics and often solely rely on a single measure, the $p$-value, to quantify the uncertainty. Utilization of Bayesian methods in scientific practice \citep{wasserstein_asa_2016} to potentially overcome the mis- and over-use of $p$-values as well as the inclusion of Bayesian methods in statistics classes at the university level \citep{berry_teaching_1997, witmer_bayes_2017, johnson_teaching_2020, hu_bayesian_2020, hoegh_why_2020, dogucu_current_2022, hu_content_2022}  have been widely recommended. 
Needless to say, the practice of scientists and their training are intertwined. 
Thus teaching students early on about dealing with uncertainty and introducing them to Bayesian ideas should be a curricular priority. 

There have been many historical changes that make Bayesian methods more popular than they used to be in the past. Perhaps the most prominent change that impacts the teaching and learning of Bayesian methods is the advances in computing. There are various tools requiring various skill levels available for teaching and learning Bayesian methods. Tools include but are not limited to point-and-click software JASP \citep{love2019jasp}, full probabilistic programming languages STAN \citep{carpenter2017stan} and JAGS \citep{plummer2003jags}, and R packages such as \texttt{rstanarm} \citep{R-rstanarm} and \texttt{tidybayes} \citep{R-tidybayes}.

Despite the availability of newer tools that make Bayesian statistics more accessible in the classroom, Bayesian courses are often geared toward students majoring in statistical, data, and mathematical sciences, and only a few are geared toward students in other STEM fields (e.g. biology, astronomy) \citep{dogucu_current_2022}. At the K-12 school level, there are also debates about introducing Bayesian ideas from different perspectives, such as interpreting and evaluating probabilities and making informal statistical inferences \citep{chernoff2014will, kazak2015bayesian, nilsson2014exploring, martignon2014proto}. We believe that making Bayesian methods in the broader STEM education community is vital and starting the teaching and learning of Bayesian thinking at earlier grade levels is also important. 

With this in mind, we developed and taught a classroom activity to support Bayesian thinking. Our goal was to design and implement the activity for pre-service mathematics and science teachers to learn about and be able to use Bayesian data analysis. An indirect goal in this was advancing learners at the grades K-12 (pre-collegiate) levels to think about and understand uncertainty through an accessible but rigorous Bayesian approach. Given the infrequent training in Bayesian methods at the college level, we prioritized training \emph{pre-service} (or, not yet teaching, in contrast to in-service teachers) mathematics and science teachers in Bayesian methods and prepare them to incorporate Bayesian ideas in their future K-12 courses. 

We first describe the relevant research in mathematics and statistics education and in science education on which our work expands, then go on to describe and discuss our teaching activity and experience.

\subsection{Relevant work in mathematics and statistics education}

The notion of \emph{uncertainty} has a unifying role in dealing with \emph{data} and \emph{chance}, the two closely related topics that are part of the mainstream school mathematics curriculum but often treated separately. With the increasing attention to developing students' informal statistical inference \citep{mr18} starting from early grades  \citep {makar2014, watson2008, ben2006scaffolding}, reasoning about uncertainty has become of interest in the context of making informal statistical inference at school level \citep{braham2015students, kazak2015bayesian, Henriques2016}. Informal statistical inference involves making claims beyond data, using data as evidence to support these claims, and using probabilistic language to make generalizations \citep{makar2009framework}. Hence, an articulation of uncertainty is at the heart of informal statistical inference. 

In the traditional method of teaching statistical inference, a common underlying reasoning process, known as the Fisherian approach, entails "assessing the strength of evidence against a claim" \citep[p.7]{rossman2008} based on the frequency interpretation of probability. An alternative form of reasoning, which is arguably more intuitive in making statistical inferences from data involves the Bayesian perspective that is based on a subjectivist notion of probability \citep{albert2002teaching}. This kind of reasoning process starts with a prior probability associated with a hypothesis or claim based on a personal judgment or experience and involves updating that probability in the light of new data \citep{rossman2008}. However, subjective probability, which is one of the main approaches to measuring probability, is not addressed in the school mathematics curricula even though there are opportunities for students to use subjective probability descriptors (impossible, less likely, more likely, and certain) for the outcomes of chance events in elementary school \citep{jones2007research}. 

Some recent research in mathematics education has focused on supporting young students' reasoning about uncertainty using subjectivist notion of probability and the Bayesian approach. In a study with 7-8-year-old children, Kazak and Leavy (\citeyear{kazak2018emergent, kazak2022emerging}) engaged them in predicting how likely a specific outcome of chance events (e.g., drawing a green jellybean from a bag) by marking it on a non-numeric happy face scale (on one end sad face for an impossible event, in the middle neutral face for an equiprobable event, and on the other end happy face for a certain event). They focused on children's personal (subjective) probability estimates and how their prior probability estimates have changed when new data were available to the children through carrying out physical experiments with 24 trials and then computer simulations with 500 or 1000 trials. The results suggested how intuitive this reasoning process could be in modifying prior probability estimates based on new evidence for even young children. 

Moreover, Kazak (\citeyear{kazak2015bayesian}) examined 10-11-year old students' reasoning about uncertainty through a task that involved assessing whether a chance game is fair or not and stating the level of confidence in their statement (conjecture) regarding the game's fairness. The game required randomly drawing one token from each of the two bags and if the two tokens are the same color, the students win the game. Students working in small groups first evaluated the fairness of the given four games based on their intuitions or personal beliefs with an explanation and marked their confidence level on a scale from 0 (not at all confidence) to 10 (totally confident). Then they played the game physically as much as they wanted and were asked to mark their confidence level about the fairness of the game again on a new scale based on the game results. In the next part, students used computer simulations to collect more data (ranging from 100 to 100000) and used information from the simulation results to update their level of confidence. The study showed that the task helped students express the subjective probability beliefs about a conjecture using the confidence scale and supported them in updating their confidence in their personal beliefs with the new information available overtime. Hence, promoting Bayesian reasoning in earlier grades of schooling appears to be promising in developing their intuitions into more powerful ideas in probability.     

Even though with age, students' perceptions of probability change \citep{piaget1951genese,kreitler1986development, barash2019heuristic}, many challenges with learning statistical inference remain even at the undergraduate level (and beyond). 
Statistical inference is often taught from the frequentist perspective and the latest recommendations include teaching inference through simulation, such as bootstrap sampling and randomization tests \citep{rossman2014using}. 
However, despite the many benefits, simulation-based activities do not always help to resolve misconceptions related to inference. For instance, in a class teaching inference with simulation, pre-service teachers had difficulties formulating hypotheses, interpreting $p$-values, and drawing conclusions \citep{biehler2015preservice}.

Although not specific to pre-service teachers, Bayesian courses and activities have also been recommended and taught at the undergraduate level as an alternative approach to statistical inference. Even though Bayesian courses are not that common at the undergraduate level \citep{dogucu_current_2022}, there are many examples of such courses \citep{witmer_bayes_2017, johnson_teaching_2020, hu_bayesian_2020, hoegh_why_2020, hu_content_2022}. Efforts have been also made to introduce Bayesian ideas through various activities, including with the aid of M\&M's candies  \citep{eadie_introducing_2019} and a web-simulator \citep{barcena_web_2019} to search for a submarine.

\subsection{Relevant work in science education}

While there has been more prior research in mathematics and statistics education that advances a Bayesian way of approaching uncertainty, there is some relevant research in science education. Two foundational papers considered the science and engineering practice---an activity common to the professional work of scientists and engineers \citep{ngss}---of \emph{arguing from evidence} in K-12 classrooms from a Bayesian perspective. 

\citet{so12} make an important distinction with respect to how and why Bayesian approaches to scientific reasoning can be useful to grades K-12 science teachers and learners. Namely, "the key leap that characterizes the debate about the value of Bayesian inference as a model of scientific reasoning" (p. 61) lies in acknowledging the degrees of belief that students hold about phenomena and scientific ideas and theories. This \emph{subjective} view of probability makes Bayes' Theorem more than a mathematical expression: instead, Bayes' Theorem can be used to understand and bolster student reasoning. They explain that this key facet of Bayesian approaches can support an informal but principled form of scientific reasoning. This focus on \emph{informal but principled scientific reasoning} is in line with similar calls in the domains of statistics education for students to have opportunities to participate in informal statistical inference \citep{mr18}.

\citet{n11} provided examples from implementing Bayesian approaches to argumentation in actual K-12 classroom contexts. While focused on argumentation, the subject matter of students' engagement in arguing from evidence centered on the question of raising taxes to provide resources to homeless individuals. Different from the emphasis on subjective probability by \cite{n11}, \citet{so12} showed how the prior and likelihood could be obtained from empirical evidence---also in the context of students engaging in scientific argumentation.

In addition to these two papers, some recent research in undergraduate physics education contexts has advanced a more \emph{qualitative} approach to Bayesian methods \citep{warren2018quantitative, warren2020impact}. Warren designed and implemented what he termed \emph{Bayesian updating activities} into introductory, university-level physics courses. In these activities, students expressed their initial confidence in the hypotheses they would test. Then, they were prompted to consider how experimental data they collected aligned with (or differed from) their initial hypotheses, after which they \emph{updated} their beliefs using Bayes' Theorem. Importantly, this work took a more qualitative approach to Bayesian reasoning: Students' initial beliefs were not expressed in terms of numbers or distributions but as hypotheses. But, this approach may still be challenging for most students at the pre-collegiate level, as applying Bayes' Theorem still requires a degree of mathematical acumen that may be out of reach for students outside of advanced high school-level mathematics or statistics courses.

\cite{rosenberg_making_2022}
extended some of this past research in science education. They explicated some epistemic (or, relating to knowledge) principles that science teachers could use with their students in a heuristic manner. These were a) \emph{be open to new evidence}, b) \emph{account for what is already known}, and c) \emph{consider alternative explanations}. In addition, they built an interactive, web-based version of the qualitative approach advanced by \citet{warren2018quantitative, warren2020impact}. But, this application does not allow the exact calculations to be made, instead using more heuristic calculations; only works for \emph{hypotheses}, rather than considering parameters. In short, it is more useful as an \emph{informal} means of introducing Bayesian reasoning. We discuss next how we move past these limitations with the application and lesson plan we developed.

\section{The Web Application and Activity Plan}

\subsection{The Web Application}

We developed a web application to assist the learning of some foundational Bayesian ideas. We wanted the users to be able to 1) specify a prior model; 2) understand that the posterior model is constructed by considering the prior model and the likelihood function (based on the data) simultaneously; and, 3) understand how  more data (i.e. evidence) causes the influence of the prior on the posterior to be weaker---and the influence of the likelihood to be stronger. 
These ideas are model-agnostic and hold true for any Bayesian model. In the web app (and the activity) we have utilized the Beta-Binomial model for its simplicity. The term \emph{simple} refers to the fact that there is a single parameter in the model and not necessarily the ease of learning. 
In addition, the Beta-Binomial model is one of the most popular models taught in Bayesian courses \citep{dogucu_current_2022}. Thus we thought it would be an appropriate model to teach mathematics and science pre-service teachers as their first Bayesian model.

The app is developed with the \texttt{R} package shiny \citep{shiny} and provides interactivity to the Beta-Binomial model visualizations in the R package in \texttt{bayesrules} \citep{dogucu_bayesrules_2021} as used in Chapters 3 and 4 of the \emph{Bayes Rules!} book \citep{johnson_bayes_2022}. The static visualization functions in this package support learning the three foundational ideas of Bayesian modeling mentioned in the previous paragraph \citep{dogucuicots}. However, in our teaching, we chose to use interactive visualizations with the Shiny framework to avoid R installation issues.
Further, since it does not require users to code, it can be used with learners who are not familiar with R. We developed the app independent of a context (e.g., coin flip) so that it can also be utilized by other educators should there be a need and interest. 

\begin{figure}[h]
\centering
\includegraphics[scale=0.3]{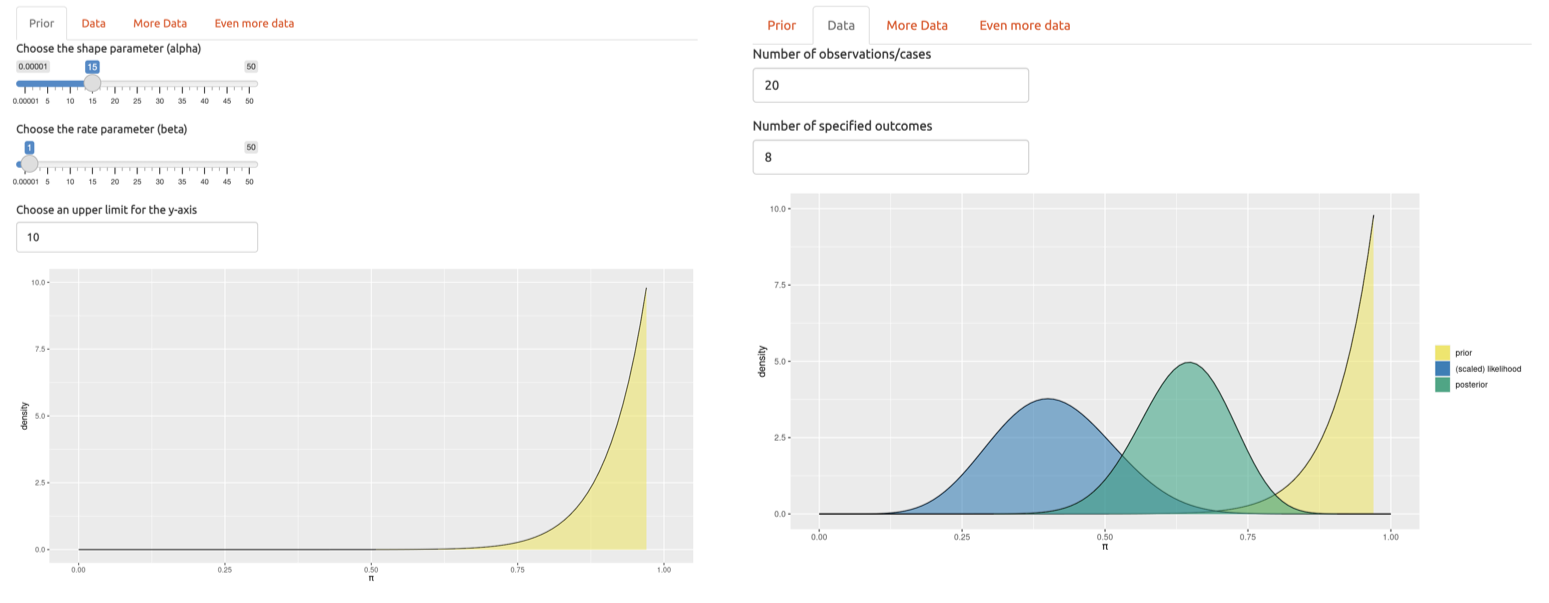}
\caption{Interface of the web app with the prior tab (on the left) and data tab(on the right).}
\label{fig:app}
\end{figure}

Using the app, the user starts by choosing the shape parameters of the Beta prior. They can change these values interactively until they find the distribution that matches their beliefs. In the next step, the user provides the data information. Once provided, then the app visualizes the prior, likelihood, and posterior. With the science educators who may not necessarily be familiar with the binomial distribution, we have avoided using terminologies such as trial and success; instead, we have used "number of observations/cases" and "number of specified outcomes," respectively. We show the first two tabs of the app in Figure \ref{fig:app}. 

The app also supports scenarios where the user can provide more data and update the posterior and even more data in a final round. These additional tabs for data and more data aim to have users visually see that with more evidence, the posterior is closer to the likelihood.

\subsection{The Activity Design}

In addition to the app, we have developed an activity that applies the Bayesian approach to analyzing and interpreting data. The aim of introducing Bayesian methods to pre-service mathematics and science teachers was two-fold. First, mathematics teachers needed to be convinced that personal (subjective) probability was useful because, with enough scientific evidence, different scientists would arrive at the same/similar conclusions, so subjective probabilities would not matter as much. The app did help seeing if the data collected have a large sample size then the  posteriors would be similar/same. Second, science teachers needed to be convinced that Bayesian thinking was useful for answering scientific questions by seeing how it could help them to analyze data in practical ways.

In order to engage students in investigating data from a Bayesian perspective, we wanted to pose a motivating problem for the students and decided to focus on a topic related to the sustainability initiatives at the university, which was energy conservation on campus. To motivate and frame the activity as an authentic one, we also wanted to pose a problem for which there was not an already-known answer. The problem involved students estimating a parameter, $\pi$, that is, the proportion of unoccupied rooms on campus with lights on. The activity consists of six parts:

\textbf{Part I - Prior Ideas:} After the introduction of the problem, students were first asked to make an initial estimate about the proportion of lights left on along with expressing how confident they were in their estimate on a scale from 0\% percent to represent being not at all confident to 100\% to represent total confidence. This estimate and the uncertainty around the estimate provided an opportunity for students to share their prior information about $\pi$. By asking them also to write the assumptions they considered when estimating the proportion, we encouraged them to present their prior observation or knowledge that gives them the basis for their initial estimate. Each student then used the applet to reflect their own prior idea about the proportion by selecting a Beta distribution determined by the shape and rate parameters (alpha and beta). The distribution that corresponded best to the student's initial idea about $\pi$ provided a prior model before collecting any data.

\textbf{Part II - Data:} Students worked in groups (4-6 students) to record data on the status of lights in unoccupied rooms from different, large buildings on campus (one per group) using a Google Sheet. 30 minutes were allocated to this data collection. The data included the assigned room name/number and whether one or more lights are on in the room (yes=1, no=0). After the data collection by groups, each student used their group's data to complete the rest of the tasks individually.  

\textbf{Part III - Change in Ideas:} In this part,  students were asked to consider only the first five observations. By using the Shiny app, students now visualized the prior, the likelihood, and the posterior distributions given their data (\emph{n}=5) and wrote down what they noticed and wondered about their prior, likelihood, and posterior distributions. They could also compare the influence of the different data and different prior on the posterior. We needed them to understand their posterior as representing an estimate that compromised between their prior ideas and data using Bayes’ Theorem. Considering the posterior, they then marked again how confident they were in their estimate on the scale (0\% to 100\%) provided. After a class discussion about taking into account the prior, the data, or both, a brief lecture on frequentist and Bayesian approaches was provided. Then, they were asked which of these approaches is more closely aligned with scientific inquiry and to explain why. 

\textbf{Part IV - More Data:} Students now used the Shiny app to analyze all of the data (\emph{n}=48-57) collected by their group. In this part of the investigation, students needed to understand that the posterior from the previous part (III) was used as the prior. Then they were asked to interpret the data and compare the updated posterior with the previous one. They marked again how confident they were in their updated estimate on the scale (0\% to 100\%) based on the new information. 

\textbf{Part V - Pooling Class Data:} In order to see the effect of a large sample size, all the data collected by the groups were combined into a whole-class data (n=205) spreadsheet. By considering the prior, likelihood of the data, and the posterior based on the whole-class data, students were asked to write about what they noticed about their estimates now and evaluated their confidence again by marking the scale given. At the end of the investigation, students drew a conclusion about the proportion of rooms with lights left on across the campus and reflected on how this investigation can be extended.

\textbf{Part VI: Reflection and Planning:} Students reflected on what they took away from the activity, what they could use with students, and what they would still like to know. Students carried out this reflection in small groups, after which they responded to prompts with short, written reflections.

\section{Implementation}

\subsection{The Instructional Context}

For the three years prior to the implementation, one of the authors taught a class for pre-service (not yet teaching) and \emph{job-embedded} (teaching while earning their teaching license) science teachers at a University in the Southeastern United States. This course focused, broadly, on project-based learning in science \citep{krajcik2018teaching}, an approach that emphasizes learning science through planning and carrying out complex projects motivated by an important scientific question or problem. 

A parallel course was taught for pre-service teachers by another instructor. In the past, the instructor of the course for pre-service mathematics teachers and the author who taught the course for pre-service science teachers combined several of their class sessions. The rationale for doing this was manifold: to support learning about the complementary subject matter (for mathematics teachers, \emph{science}, and for science teachers, \emph{mathematics}) and to engage the pre-service teachers in these classes in activities that integrate mathematics and science - namely, \emph{data modeling} \citep{kazak2021students, lehrer2018introducing}, as work with empirical data represents a relatively rare area of overlap between the science and mathematics curricular standards (the Next Generation Science Standards and the Common Core State Standards, both of which were adopted for the standard's used in the state). Findings from these earlier efforts are documented in \citet{lawson2021better}. Because of these earlier attempts to work with both mathematics and science pre-service teachers around data, this was a suitable context for exploring the Bayesian approach that we advance and describe in this paper. Both instructors, thus, were experienced with teaching these classes, and both were interested in supporting their students in these classes - current and future teachers - to be able to engage their (grades K-12/pre-collegiate) students to work with data in more ambitious, meaningful ways. 

\subsection{Participants}

As described in the previous section, the sample for this work involved pre-service (earning their teaching license) science and mathematics teachers. We note that a few students were technically \emph{job-embedded} (concurrently teaching and earning their teaching license) teachers. Though \emph{University-level students} in the course we taught, we refer to them in this manuscript as \emph{pre-service teachers} as this was their primary identity in the course. We combined the two classes for a single, approximately three-hour class session during the Fall 2021 semester. Notably, the COVID-19 pandemic was a severe challenge during this semester, and, accordingly, both of the two classes were offered in a hybrid modality: pre-service teachers could join online if they were concerned about (or exposed to) COVID-19, and many did based on their preferences that changed on a week-by-week basis. More specifically, there were 21 science and 14 mathematics pre-service teachers enrolled in the classes.  The pre-service teachers had limited prior exposure to and experience with statistics, data analysis, and data science. None of the (middle or high school) pre-service science teachers' programs of study required any statistics or data science courses. High school pre-service mathematics teachers' program of study required a course on probability and statistics (with Calculus 3 as a pre-requisite, meaning that few of the participants in this study had taken this course at this point in their program).

Of the 35 pre-service teachers, practically all participated in the activity. However, nine science and seven mathematics pre-service teachers consented for us to use their work products as data sources for this study, for a total of 16 participating pre-service teachers. Three of those pre-service teachers did not share their complete work with us, and so our analytic sample consisted of 13 pre-service teachers---seven science and six mathematics. 

\subsection{Accommodations}

As noted earlier, the classes were offered in a hybrid modality; for the class session in which we carried out the Bayesian data analysis activity, several pre-service teachers joined in a fully online modality, and we took several steps to accommodate these pre-service teachers with the aim of ensuring that they had opportunities to fully engage in all class activities. Specifically, we constructed groups of pre-service teachers that not only combined pre-service teachers in the mathematics and science classes (so that they had opportunities to network and collaborate with pre-service teachers teaching a different subject area) but also pre-service teachers joining face-to-face and online. Because pre-service teachers in the online modality typically joined for around two (instead of the full three hours) of the classes in earlier sessions, we shared all materials ahead of time and recorded a video for online pre-service teachers to view in advance of the class. We instructed pre-service teachers in the face-to-face classroom to advocate for the online teachers in their group and to be responsible for their online groupmates to be able to participate; accordingly, teachers in the face-to-face noted to the instructors when their online groupmates had challenges hearing audio or when they had questions to raise to the whole class. While pre-service teachers worked in groups and talked through all aspects of the task, they completed and submitted the task independently.

\subsection{Materials}

The materials used consisted of the following (see the Appendices for links to these materials):

\begin{enumerate}
    \item \emph{The Shiny interactive web application.} The application enabled pre-service teachers to estimate a posterior using a beta-binomial model that we described earlier.
    \item \emph{A brief Google Form for pre-service teachers to self-assess their degree of Bayesian and Frequentist thinking}. This consisted of four questions that teachers could answer and then interpret the meaning of in terms of how much their thinking aligned with a Bayesian or Frequentist approach to data analysis (from the Bayes Rules! book \citep{dogucu_bayesrules_2021}).
    \item \emph{A Google Docs document that served as a digital worksheet}. This document consisted of six parts that aligned with how we structured the activity for pre-service teachers: 1) prior ideas, 2) data, 3) change in ideas, 4) more data, 5) pooling class data, and 6) reflection. 
    \item \emph{Google Sheets spreadsheet for pre-service teachers to record observations as data}. This spreadsheet included separate tabs for teachers to record their data in groups---and a tab to pool the entire class's data. Summaries of these data were used in the app.
    \item \emph{A Google Slides presentation}. This was used to introduce new ideas to pre-service teachers and to help organize the task.
\end{enumerate}

\subsection{Participant comments}

To provide content and formative feedback on the implementation, we collected and analyzed teacher comments in the form of their responses in the Google Docs document. 

The key points taken away by the pre-service mathematics and science pre-service teachers after completing this activity were mainly related to the content, lesson format, and technology. When the pre-service teachers commented on their experiences with the content aspect of the activity (i.e., Bayesian data analysis), they tended to consider the concepts and approaches used to analyze and interpret data. Some examples of such comments are "I now have a better intuition regarding the differences between prior, likelihood, and posterior probabilities." and "That there are many different methods to interpret and draw conclusions from data.". 

The pre-service teachers' comments about the lesson format (i.e., "interactive lessons" and "group activity") showed their predominant orientation towards teaching. For instance, as seen in the comment "This is a great group activity for teachers, for our future teachers, to be introduced to the value and limitations of data.", the pre-service teacher appeared to reflect on his experience and relate it to his teaching in the future. Moreover, some pre-service teachers mentioned learning how to use the Shiny app as a new tool to analyze the data. 

The activity seemed to also foster pre-service teachers' interest in Bayesian ideas (and, more generally, statistics and probability), which purportedly were not part of their teacher training, per their responses. For instance, a pre-service (mathematics) teacher  stated, "I want to learn more about Bayes’ Theorem and how it can be applied in other situations in the real world. I also want to learn about other ways of representing data. I have never been in a Statistics class, but this was very interesting." Some pre-service teachers (especially those preparing to teach science) were interested in learning more about the app and visualizing data for their own teaching (e.g., "How I can make more data visual for my students").   

When the pre-service teachers were asked to reflect on other topics that they could use a similar approach with their students, they mostly gave general data collection examples in different contexts, such as measurement, functional relationships, bird surveys, and even quantum physics. Hence, the perception that the Bayesian approach can be used in any data analysis seemed to be common among the pre-service teachers. This range was reflected in specific examples of ways that the Bayesian approach could be used that the pre-service teachers noted. Particularly, while a pre-service mathematics teacher provided a binomial example (i.e., a survey about what people like or do not like) similar to the activity, a pre-service science teacher related the approach used in the activity to his subject area, such as testing hypotheses ("You could use the app to analyze data that students collecting in an experiment and compare their hypothesis (prediction) to the results.").

\section{Conclusion and Discussion}

Our aim was to design and implement a Bayesian data analysis activity for pre-service mathematics and science teachers. We reasoned that this would show that it is possible for pre-service science teachers with no background in Bayesian statistical methods to do such an activity. In this section, we discuss the activity and our observations from teaching to pre-service teachers. We do this threefold. First, we discuss Bayesian thinking and its place in teacher education and how this activity contributes to the preparation of science and mathematics teachers. Second, we revisit our indirect aim of bringing Bayesian thinking to the K-12 level. Last but not least, the participants we worked with are college students in addition to being pre-service teachers. Thus we discuss the activity and its contributions from college-level statistics education perspective.

\subsection{Contributions to Teacher Education}

One contribution this work makes is demonstrating the viability and value of bringing together pre-service mathematics and science educators. Traditionally, pre-service teacher preparation has held teachers in these two content areas largely separate---though there is one prominent program that integrates the coursework of future mathematics and science teachers \citep{backes2018can}. Still, mostly, mathematics teachers learn to teach mathematics with other mathematics teachers, and science teachers with other science teachers. Helping students to analyze data is a hallmark of both the mathematics and science standards \citep{ccss, ngss} and it is therefore not only viable in terms of curricular standards to design and implement course experiences for mathematics and science educators, but also possibly valuable because of the complementary differences in mathematical and science approaches to data analysis. Namely, mathematical modeling has historically been criticized as ignoring the \emph{context} in which data analysis occurred \citep{rubin2020}. At the same time, how science teachers introduce data to their students may lack statistical rigor \citep{rosenberg2022big}. This became apparent in the participants' reflections at the end of the activity when two mathematics pre-service teachers commented that they observed mathematics and science pre-service teachers had different approaches to statistical concepts and data analysis. A Bayesian data analysis activity and lesson like the one we implemented was designed to balance these two historical deficiencies in the professional preparation of teachers---combining a meaningful \emph{context} with mathematical \emph{rigor}. We observed teachers helping one another with the aspects of the activity and lesson with which they had less experience: mathematics teachers commented to science teachers about how the distributions they observed in the map connect to calculus-related ideas, and science teachers commented to mathematics teachers about specific environmental science and physics-related ideas. Such opportunities can therefore make the other content areas more accessible to both groups of teachers while still meeting the curricular standards because of the central role of data in each.

In this activity, we tried to make transparent/concrete how one's initial hypothesis based on a personal judgment can be updated with the availability of new data in a real-life context relevant to college students. The pre-service teachers started by stating their initial prediction and their confidence level in their estimate. Then, they analyzed the prior-likelihood-posterior distributions in the app and updated their confidence level with the first five data collected by their group (\emph{n}=5), with all group data (\emph{n}=48-57), and finally with the class data (\emph{n}=205). This approach relied on the intuitiveness of the deductive reasoning used in the Bayesian approach, as emphasized by others \citep{albert2002teaching, rossman2008}, and made this activity potentially useful for pre-service teachers even without prior undergraduate coursework in probability, statistics, and calculus. Moreover, some pre-service teachers' comments showed an interest in learning statistics, probability---and more specifically---Bayesian ideas after completing this activity in a three-hour class. Hence, such activities can be incorporated into a teacher education course that is relevant to the Bayesian data analysis content, such as the course focusing on project-based learning in science used in this study. 

Another contribution this study makes to teacher education is a demonstration of a technology specifically designed for teaching and learning. In teacher education, there is a longstanding interest in how teachers use content- (e.g., mathematics or science education) and context-specific (e.g., usable given the technologies at hand in K-12 schools) technologies \citep{mishra2006}. The app we designed had these considerations in mind. It is not a tool designed (or necessarily useful) for Bayesian data analysis \emph{in general}; other widely-used tools---such as Stan \citep{gelman2015}---exist for that. At the same time, the tool was designed to enable learners to carry out analyses that are statistically and technically valid. Moreover, it is designed to be able to be usable by any student and teacher via a computer (or a Chromebook) with a web browser. For these reasons, this study contributes an example of statistical software for teaching and learning that is intended to partially address the call of \citet{mcnamara2019} to design tools that bridge the gap between those strictly for learning and those for professionals. The tool we developed is available for any teacher to use. We can envision extensions of this tool that permit analyses using different distributions for the dependent variable (i.e., the normal-normal and Gamma-Poisson conjugate prior models) that further bridge between what teachers and learners could do in K-12 classrooms using the app we developed and what tools such as Stan permit more sophisticated users to do. 

\subsection{Contributions to K-12 Education}

 There have been emerging opportunities for introducing Bayesian ideas at the school level in probability and informal statistical inference contexts \citep{chernoff2014will,kazak2015bayesian,nilsson2014exploring,martignon2014proto}. This study complements those works as we indirectly intended for fostering students' understanding about uncertainty through the Bayesian approach in K-12 levels by implementing our Bayesian data analysis activity with pre-service mathematics and science teachers. After completing the activity and lesson, the pre-service teachers' comments indicate their intention to use several aspects of the activity, such as lesson format (group activity), content (the value and limitation of data), and the app, to foster student learning in their own classrooms.

A research-related contribution this work makes to K-12 science education is an example of how Bayesian data analysis applies in this context. There is prior research on how Bayes' Theorem can be used in science classrooms, but this work has explored a Bayesian approach to what is commonly referred to in science education research as the \emph{science and engineering practice} of argumentation \citep{n11, so12}. There are a few examples of Bayesian data analysis, but this work has been conducted in the undergraduate physics context \citep{warren2018quantitative, warren2020impact}. Thus, the activity and lesson we described in this article present the first example of which we are aware of how teachers at the K-12 level can support their students to analyze data in a Bayesian manner. We hope that future research---ours or others---makes further contributions in this area by beginning to document the impacts of engagement in Bayesian data analysis using the kinds (or adaptations) of research design and assessments used in the work of \citep{warren2020impact}.

Another contribution of this study to K-12 education is to describe how to implement a Bayesian approach to reasoning about uncertainty that provides a natural way to use additional data to update prior probability estimates and beliefs about confidence for future mathematics and science teachers. There are three main approaches to probability measurements: classical (known as 'theoretical' in school mathematics), frequentist (known as 'experimental' in school mathematics), and subjective. While some aspects of the first two approaches are incorporated in K-12 mathematics curricula in different countries, there is a lack of  treatment of the subjective probability in these documents \citep{jones2007research}. As also noted by Jones et al. and supported by more recent research \citep{kazak2015bayesian, kazak2022emerging}, there is some evidence that even young students can develop intuitions about modifying prior probability estimates or beliefs about confidence levels when new data are available in engaging with subjective probability. So, pre-service mathematics and science teachers who have such learning experiences themselves will be more likely to use these ideas to help their students develop the kind of knowledge needed to make personal probability judgments.    

The Bayesian Data Analysis activity presented in this paper involves ideas related to informal statistical inference \citep{makar2009framework}  and uncertainty. These ideas are potentially relevant to both mathematics and science curricula in K-12 level as analyzing and interpreting data with an articulation of uncertainty are part of the mathematics and science standards (see the Common Core State Standards: Mathematics (CCSSM) http://www.corestandards.org/Math and the Next Generation Science Standards https://www.nextgenscience.org/). Even though mathematics and science contents are taught in isolation at school level, mathematics can provide the foundations for analyzing data in solving the real-world problems and science can be a rich source of meaningful contexts for data investigations \citep{watson2017linking}. We anticipate that the implementation of this activity within a scientific context (sustainability/energy conservation) with the participation of both mathematics and science pre-service teachers can inspire efforts to make connections between mathematics and science contents in the K-12 school curriculum.



    
\subsection{Contributions to College Statistics Education}

We have mentioned earlier in Section \ref{sec:intro} that the statistics community calls for scientists to consider Bayesian methods as an alternative in data analysis \citep{wasserstein_asa_2016} and the statistics education literature shows evidence of courses at the college level offered in statistics departments \citep{berry_teaching_1997, witmer_bayes_2017, johnson_teaching_2020, hu_bayesian_2020, hoegh_why_2020, hu_content_2022} but these courses do not seem to be in the broader STEM programs \citep{dogucu_current_2022} with very few exceptions. The fact that Bayesian courses have many prerequisites \citep{dogucu_current_2022} could be one reason that these courses are not accessible to the broader STEM community. 

One prerequisite that is often the centerpiece of discussions on statistical training is calculus. When it comes to introduction to statistics courses (which often are taught in the frequentist paradigm), in the modern-day curricula, they are taught with or without calculus as a prerequisite \citep{gaise_guidelines_2016}. Even though, Bayesian statistics courses rely heavier on calculus as a prerequisite, the fundamentals of Bayesian reasoning, such as updating beliefs, can even be communicated to young learners \citep{kazak2015bayesian}. We believe that this activity adds to the body of literature on Bayesian activities (e.g. \citep{eadie_introducing_2019, barcena_web_2019}) that can be completed with high school algebra knowledge. The activity is intentionally designed to avoid mathematical derivations and focuses on fundamentals of Bayesian thinking with the aid of visualization similar to examples in \cite{dogucuicots}. Thus it can serve to introduce Bayesian concepts to college-level students who meet fewer mathematics and statistics prerequisites and cannot access a full course on Bayesian statistics.

\subsection{Limitations}

Though we think this activity makes several contributions, we also note some limitations. This was a single activity with two classes and we think it will be fruitful to expand the group of teachers to understand how well this activity meets the needs of teachers at different grade levels, with differing degrees of mathematical and scientific knowledge, and in different teaching contexts (e.g., schools with different levels of socioeconomic capital). We also think it will be fruitful to expand to teachers at different career stages---namely, to \emph{in-service} (independently teaching in their classroom ) mathematics and science teachers.

Furthermore, we think that expanding the activity to include multiple classes---a "unit" in a course or even an entire course---could be beneficial, though we do think the initial exposure we provided was still valuable. 

Another limitation we would like to consider addressing in future iterations is balancing necessary background information with the core activity that pre-service teachers completed. Given the duration of the single class, a great deal of information was provided to pre-service teachers about Bayes' Theorem, Bayesian data analysis methods, and the use of the app. Some of this material could be provided in advance to allow more time for in-class work, discussion, reflection, and planning by teachers.

A last limitation concerns the degree of emphasis on the scientific ideas relative to the emphasis on the Bayesian approach. In particular, we think it would be beneficial to align this activity with the disciplinary core ideas that comprise the curricular standards in the country in which we implemented the activity (i.e., the Next Generation Science Standards; \citep{ngss2013}. We think this would involve identifying a scientific phenomenon that has a mechanism or process that can be readily represented with the model type (Beta-Binomial) we used. Or, we could use a different appropriate model type (e.g., Normal-Normal or Gamma-Poisson). Doing so would heighten the stakes for understanding the scientific focus of the activity while retaining the emphasis on understanding how Bayesian methods proceed.

The learning activity and the accompanying web app can be used as is by anyone teaching pre-service, in-service teachers, or equally trained learners. Since these resources are provided as open-source, alterations can be made to adopt these resources in different contexts. Based on our experience in teaching this activity and some limitations that we faced, we suggest the following as possible alterations that the readers may consider:

- If teaching only mathematics pre-service or in-service teachers then the activities can be supported by a further in-depth mathematical explanation of the Beta-Binomial model. Many of our participants who are pre-service mathematics teachers were eager to learn why the model worked the way it did.

- If desired, the scientific context can possibly be changed. In this case, readers can choose any other unknown to represent $\pi$. This has to be done with caution as the data are assumed to follow a Binomial model which assumes that each observation is independent of the other.

- Bring together STEM teachers. A Bayesian modeling activity is a perfect opportunity to bring together teachers of different subjects. It is worth noting though this can be logistically difficult. 

- Dedicate time for data collection outside of class time. Unsurprisingly, teaching time is never enough whether it is teaching Bayesian modeling or any other topic. Teachers, especially those who are less familiar with data modeling, need time to internalize concepts. One alteration that future adaptations can include is that teachers can collect data outside of class time thus freeing more time for clarifying concepts during class time. Collecting data at different times can also help with the generalizability of the findings.

- Alter activities based on the mode of teaching (i.e. hybrid, online, or in-person). Future adaptations should carefully take into consideration the data collection process. For instance, if the activity were taught online, the data scenario needs to be changed so that teachers would be able to collect the data at home and in their dormitories. 

- Last, consider using different tools. For instructors who would like to adopt a different Bayesian model but is not interested in / able to utilize the Shiny package to develop an app of their own, JASP \citep{love2019jasp} may be appropriate. 


\section{Closing Remarks}

Bayesian data analysis is potentially useful and empowering to students at the grades K-12 (pre-collegiate) levels. Teachers are an essential part of helping their own students to potentially benefit from such an approach. In this work, we designed and developed a lesson for pre-service teachers that teachers experienced and reflected on. In so doing, we showed that it is readily possible for pre-service teachers to use such an approach and to report immediate and potentially longer-term benefits from doing so. We hope this work instigates future research and design and development that is intended to make Bayesian thinking more accessible but as powerful for grades K-12 learners as it is in the many domains in which it is used.

\section*{Appendices}

\begin{enumerate}

    \item The web app is freely accessible online at \shinyapp{} and the source code is available on GitHub \shinyrepo{}. 
    \item The (Electronic) Handout provided to pre-service teachers during the activity can be accessed at \googledoc{}
    \item Slides presented to pre-service teachers during the activity can be found at \slides{}
\end{enumerate}

\section*{Data Availability}

The data that support the findings of this study are available from the corresponding author, MD, upon reasonable request.	

\section*{Human Participants}

Informed consent was obtained from all study participants. This study was approved by the \irb{} Institutional Review Board (18-04804-XP).

\bibliographystyle{agsm}

\bibliography{bibliography, bayesed}

\end{document}